\documentclass[preprint,showpacs,preprintnumbers,amsmath,amssymb,nofootinbib]{revtex4}
\usepackage{amsfonts}
\usepackage{booktabs}
\usepackage{mathrsfs}
\usepackage{epsfig}
\usepackage{graphicx}
\usepackage{dcolumn}
\usepackage{bm}
\usepackage{amsmath}
\usepackage{slashed}

\let\jnfont=\rm
\def\NPB#1,{{\jnfont Nucl.\ Phys.\ B }{\bf #1},}
\def\PLB#1,{{\jnfont Phys.\ Lett.\ B }{\bf #1},}
\def\EPJC#1,{{\jnfont Eur.\ Phys.\ Jour.\ C }{\bf #1},}
\def\PRD#1,{{\jnfont Phys.\ Rev.\ D }{\bf #1},}
\def\PRL#1,{{\jnfont Phys.\ Rev.\ Lett.\ }{\bf #1},}
\def\MPLA#1,{{\jnfont Mod.\ Phys.\ Lett.\ A }{\bf #1},}
\def\JPG#1,{{\jnfont J.\ Phys.\ G}{\bf #1},}
\def\CTP#1,{{\jnfont Commun.\ Theor.\ Phys.\ }{\bf #1},}
\def\ZPC#1,{{\jnfont Z.\ Phys.\ C }{\bf #1},}
\def\JHEP#1,{{\jnfont JHEP \ }{\bf #1},}
\def\Rv{\not{\hbox{\kern-1pt $R$}}}
\def\p{\not{\hbox{\kern-3pt $p$}}}

\begin{document}
\preprint{\parbox{1.2in}{\noindent arXiv:1109.6543}}

\title{Testing new physics models
       by top charge asymmetry and
       polarization at the LHC}

\author{Junjie Cao$^1$, Ken-ichi Hikasa$^2$, Lin Wang$^1$, Lei Wu$^3$, Jin Min Yang$^3$
        \\~ \vspace*{-0.3cm} }
\affiliation{ $^1$ Physics Department, Henan Normal University,
     Xinxiang 453007, China\\
$^2$ Department of Physics, Tohoku University, Sendai 980-8578, Japan \\
$^3$ Institute of Theoretical Physics,
     Academia Sinica, Beijing 100190, China
     \vspace*{1.5cm}}

\begin{abstract}
As a top quark factory, the LHC can test new physics models used
to explain the top quark forward-backward asymmetry $A^t_{\rm FB}$
measured at the Tevatron. In this work we perform a comparative
study for two such models: the $W^\prime$ model and the color
triplet diquark ($\phi$) model. Requiring these models to explain
$A^t_{\rm FB}$ and also satisfy the top pair production rate
measured at the Tevatron, we examine their contributions to the LHC
observables such as the polarizations and charge asymmetries in
top quark productions and the charge asymmetry in $W^\prime$ (or
$\phi$) pair production. We find that these observables can be
enhanced to their observable levels and current LHC measurement
on the top charge asymmetry has already tightly constrained the
$W^\prime$ model. We also find that each observable shows different
characteristics in different models, which can be utilized to
discriminate the models.
\end{abstract}
\maketitle

\section{INTRODUCTION}
So far the top quark properties measured at the Tevatron are in good
agreement with the Standard Model (SM) predictions except the
inclusive\footnote{We do not consider the CDF 3.4$\sigma$
discrepancy of $A^t_{FB}$ for $m_{t\bar{t}}>450$ GeV because it
is not confirmed by the D0 collaboration.} forward-backward
asymmetry $A^t_{\rm FB}$ \cite{afb-sm}, which, as reported by the
CDF collaboration and the D0 collaboration, exceeds the SM
prediction by about $2 \sigma$ \cite{top-afb-cdf,top-afb-d0}. Such
an anomaly has been widely speculated as a harbinger of new physics
and thus stimulated various explanations in extensions of the SM
\cite{afb-review,axigluon,afb-s-channel,light-zprime,non-abel,Asy-LR,afb-tu-channel,afb-eft}.
These extensions, albeit in quite different forms, usually have rich
top quark phenomenology at colliders. Since the Tevatron is going to
be shut down very soon, the task to screen out the right theory is
left for the LHC \cite{top-review}.

Although the present top quark dataset at the LHC is moderate, it is
already capable of scrutinizing the validity of some extensions. For
example, the non-observation of a clear resonance in the $t\bar{t}$
production searched by the ATLAS and CMS Collaborations at
$\sqrt{s} = 7$ TeV implies that an axigluon with strong couplings
to light quarks should be heavier than 3.2 TeV \cite{resonant-lhc},
which makes it less attractive as an explanation of $A^t_{\rm FB}$
\cite{axigluon} ( however, as pointed in the last reference in
\cite{axigluon}, a light axigluon with an enlarged width and reduced
couplings to light quarks is still allowed by the current LHC
measurements). Meanwhile, since no excess of same-sign top quark
events was observed by recent measurements from the LHC and
Tevatron \cite{like-sign-top,like-sign-top-exp}, the light
$Z^\prime$ model based on flavor non-universal $U(1)$ symmetry
\cite{light-zprime} is also disfavored. Among the surviving models
two typical ones are the $W^\prime$ model \cite{wp} and the
diquark ($\phi$) model \cite{triplet}, which, as pointed in
\cite{afb-shu}, are preferred by the combined fit of $A^t_{\rm
FB}$ and the total $t\bar{t}$ production rate measured at the
Tevatron. In this work we focus on these two models and perform a
comparative study by considering several observables at the LHC. Our
study shows that each of these observables can be enhanced to the
observable level and meanwhile exhibits different characteristics in
these two models. As a result, the $W^\prime$ model is found to be
tightly constrained by the charge asymmetry in $t\bar{t}$ production
at the LHC, while the diquark model can be readily explored once
more luminosity is accumulated at the LHC.

We will consider the following observables:
\begin{itemize}

\item[(i)] Top quark charge asymmetry in $t\bar{t}$ production at the LHC,
which is defined by \cite{cms-ac}
\begin{eqnarray}
A_{C}(t\bar{t})=\frac{\sigma(|\eta_{t}|>|\eta_{\bar{t}}|)-\sigma(|\eta_{t}|
<|\eta_{\bar{t}}|)}{\sigma(|\eta_{t}|>|\eta_{\bar{t}}|)+\sigma(|\eta_{t}|<
|\eta_{\bar{t}}|)}, \label{define-AC}
\end{eqnarray}
where $\eta_t$ ($\eta_{\bar{t}}$) is the pseudo-rapidity of the top
(anti-top) quark in the laboratory frame, and $\sigma$ denotes cross
section. This asymmetry reflects whether the top quarks on average
are more boosted than the anti-top quarks or not. We note that the
CMS Collaboration has recently measured this quantity with an
integrated luminosity of 1.09 fb$^{-1}$ and obtained $A^{\rm
exp}_{C}(t\bar{t}) =-0.016\pm0.030({\rm
stat.})^{+0.010}_{-0.019}({\rm syst.})$, which is consistent with
its SM prediction $A^{\rm SM}_{C}(t\bar{t})=0.0130(11)$
\cite{cms-ac}.  A similar result is also reported by the ATLAS
Collaboration with larger uncertainties \cite{atlas-ac}. So this
asymmetry can be used to limit new physics models
\cite{ac-lhc-th1,ac-lhc-th2}.

\item[(ii)] Top quark polarization asymmetry in $t\bar{t}$ production at
the LHC, defined by \cite{pt-sm}
\begin{eqnarray}
P_{t}=\frac{(\sigma_{+-}+\sigma_{++})-(\sigma_{-+}+\sigma_{--})}{\sigma_{+-}
+\sigma_{++}+\sigma_{--}+\sigma_{-+}}
\end{eqnarray}
with the first (second) subscript of $\sigma$ denoting the helicity
of the top (anti-top) quark. Unlike light quarks, top quark decays
rapidly before forming any hadronic bound state. So its spin
information is preserved by its decay products and can be recovered
by their angular distributions. For the $t\bar t$ production at the
LHC, the top quark is not polarized at the leading order of the SM
because the production proceeds mainly through the QCD interaction
and the parity-violating electro-weak contribution to the
polarization is negligibly small \cite{pt-sm}, but any addition of
new parity-violating interaction of top quark may induce sizable
polarization asymmetry \cite{pt-np1,pt-np2,afb-pt}.

\item[(iii)] Enhancement factor of the $t\bar{t}$ production rate
in  high invariant mass region of $t\bar{t}$:
\begin{eqnarray}
R_{1}=\sigma_{\rm tot}(M_{t\bar{t}}>1\,{\rm TeV})/
\sigma_{\rm SM}(M_{t\bar{t}}>1\,{\rm TeV}),  \label{R1}
\end{eqnarray}
where $\sigma_{\rm tot}$ incorporates the contributions from the SM
and the new physics. In exotic $t$-channel or $u$-channel $t\bar{t}$
production, the Rutherford singularity can alter significantly the
distribution of the $t\bar{t}$ invariant mass in high energy tail
\cite{inv-mass-lhc}, so $R_1$ may deviate significantly from unity.

\item[(iv)] Charge asymmetry in the associated production of a single
top with a particle $X$:
\begin{eqnarray}
R_{2}&=&\sigma(t X^{-})/\sigma(\bar{t} X^{+}).  \label{R2}
\end{eqnarray}
This asymmetry can be measured by requiring that the top quark decay
semi-leptonically and $X$ decay hadronically, and looking for the
asymmetry between the event numbers with one lepton and one
anti-lepton in the signal respectively. It was once suggested in
searching for single top production in the SM and in limiting new
physics models \cite{single-top-ac1,single-top-ac2}. Depending on
$m_X$ and the initial partons in $tX^{\pm}$ production, $R_2$
may be far larger or smaller than unity.

\item[(v)] Charge asymmetry in $X^+X^-$ production defined by
\begin{eqnarray}
A_{C}(X^+X^-)=\frac{\sigma(|\eta_{X^{-}}|>|\eta_{X^{+}}|)-\sigma(|\eta_{X^{-}}|
<|\eta_{X^{+}}|)}{\sigma(|\eta_{X^{-}}|>|\eta_{X^{+}}|)+\sigma(|\eta_{X^{-}}|
<|\eta_{X^{+}}|)},
\end{eqnarray}
Like $A_{C}(t\bar{t})$, this asymmetry reflects whether $X^-$ or
$X^+$ is more boosted. Given the interactions of the particle $X$
with quarks, this asymmetry is determined by $m_X$ and the
energy of the LHC.
\end{itemize}

This paper is organized as follows. In Sec. II, we briefly describe
the features of the $W^\prime$ model and diquark model. Then in
Sec. III we discuss some observables in
$t\bar{t}$ production, single top production and $W'$ ($\phi$)
pair production. Finally, we draw our conclusion in Sec. IV.

\section{The $W^\prime$ model and the diquark model}
Among various explanations of the $A^t_{\rm FB}$ anomaly, the model
with a color singlet $W^\prime$ is a promising one
\cite{wp,afb-shu}. This model can be realized in an asymmetric
left-right framework \cite{Asy-LR,mfv} presented in Appendix A,
which is based on the gauge group $SU(2)_L
\bigotimes SU(2)_R \bigotimes U^\prime (1)$ and
assumes that only the first and third generation right-handed quarks
transform nontrivially under the group $SU(2)_R$. The
interaction relevant to our calculation is given as
\begin{eqnarray}
\mathcal {L} &=&-g_{R} \bar{t}\gamma^{\mu} P_{R} d W^{\prime+}_{\mu}
+ \hbox{h.c.}\;. \label{wprime}
\end{eqnarray}
The $t\bar{t}$ production then gets additional contribution from the
$t$-channel process $d\bar{d} \to t \bar{t}$ via exchanging a
$W^\prime$, which may sizably alter $A^t_{\rm FB}$ at the Tevatron.
Note that in the framework presented in Appendix A, besides
$W^\prime$, the newly predicted  neutral and charged Higgs bosons
can also contribute to the $t\bar{t}$ production. Since the size of
such contribution is model-dependent and may be negligible if these
fields are heavy and/or the vev of $\phi_R$ is much higher than the
electro-weak breaking scale \cite{Asy-LR,mfv}, we in our study do
not consider these contributions.

Another model we are considering is the color-triplet diquark model
\cite{triplet}, where a new scalar $\phi$ (called diquark) is
assigned with the quantum number ($\bar{\textbf{3}}$, $\textbf{1}$,
$-4/3$) under the SM gauge group ${\rm SU}(3)_C\times {\rm SU}(2)_{L}
\times {\rm U}(1)_{Y}$.
The relevant Lagrangian is then given by
\begin{eqnarray}
\mathcal {L}&=&
D_{\mu}\phi^{\dagger}D^{\mu}\phi-M^{2}_{\phi}|\phi|^{2}+f_{ij}
\bar{u}_{i\alpha}P_{L}u^{c}_{j\beta}\epsilon^{\alpha\beta\gamma}
\phi^{\dagger}_{\gamma} + \hbox{h.c.}\;,  \label{phi}
\end{eqnarray}
where the coupling coefficients satisfy $f_{ij}=-f_{ji}$ with $i,j$
being the flavor index, $\epsilon^{\alpha\beta\gamma}$ is the
antisymmetric tensor in color space, and $u^{c}=C\bar{u}^{T}$ with
$C$ being the charge conjugate matrix. In this framework, the
discrepancy of $A^t_{\rm FB}$ can be alleviated by the
contribution of the $u$-channel process $u\bar{u}\rightarrow
t\bar{t}$ mediated by the triplet $\phi$. In \cite{diquark}, a
comparative study of $A^t_{\rm FB}$ was performed in diquark
models where $\phi$ is assigned in different representations of the
$SU(3)$ group, and it was found that the triplet model is better
suited to explain the $A^t_{\rm FB}$ anomaly without conflicting
with other experimental results. In our analysis, in order to escape
constraints from low energy processes such as $D^0$--$\bar{D}^0$
mixing, we set $f_{ij}$ to be zero except $f_{ut}$.

The common feature of the two models comes from the calculation of
the $t\bar{t}$ production rate, where the interference of the new
contribution with the SM QCD amplitude always partially cancels
the pure new contribution. In fact, this cancellation is essential
for the models to explain the $A^t_{\rm FB}$ anomaly and at same
time keeps other observables consistent with their measured values
at the Tevatron. We checked that such cancellation persists in
calculating $A_C$ discussed below, and the extent of the
cancellation depends on the new particle mass and the collider
energy. We also checked that, partially due to the difference in
parton distributions for the initial states, $A_{\rm FB}^{t}$ in
the diquark model usually exceeds that in the $W^\prime$ model if
$g_R = f_{ut}$ and $m_{W^\prime}=m_{\phi}$.

\section{ Numerical results and discussions}
In this section we present the numerical results for the observables at the LHC with
$\sqrt{s} = 7$ TeV.  We take the SM parameters as \cite{pdg}
\begin{eqnarray}
m_t=172.5{\rm ~GeV},~m_{Z}=91.19 {\rm
~GeV},~\sin^{2}\theta_W=0.2228. ~\alpha_s(m_t)=0.1095,~\alpha=1/128,
\end{eqnarray}
and use the parton distribution function CTEQ6L1 \cite{cteq} by
setting $\mu_R=\mu_F$ with $\mu_R$ and $\mu_F$ denoting the
renormalization scale and the factorization scale respectively.

For the constraints from the $t\bar{t}$ production rates, we
consider the Tevatron measurements \cite{top-cross-exp-tev}, which
are so far the most precise results\footnote{The latest LHC
measurement  \cite{top-cross-exp-lhc} has marginally reached the
Tevatron precision. If we consider the LHC limits, our results
remain unchanged.}. We require the predictions of the inclusive
$A^t_{\rm FB}$ and the total $t\bar{t}$ production rate in each
model to lie within $1\sigma$ region of their experimental values.
As mentioned earlier, we do not consider the discrepancy of the
$A^t_{\rm FB}$ in large $t\bar{t}$ invariant mass region reported by
the CDF collaboration (about $3.4\sigma$ away from its SM prediction
for $M_{t\bar{t}} > 450$ GeV\cite{top-afb-cdf}) since it is not
confirmed by the D0 collaboration \cite{top-afb-d0}. We also do not
consider the constraint from the measured $t\bar{t}$ invariant mass
distribution at the Tevatron because the shape of such a distribution
in high energy tail is sensitive to the cut efficiency of event
selection and also to QCD corrections \cite{afb-shu,non-abel}.

\subsection{Observables in $t\bar{t}$ production}
Before presenting our results for $A_C(t\bar{t})$, we point out
two features of $A^t_{\rm FB}$. First, because the valence quark
in proton always moves in parallel with the proton, $A^t_{\rm FB}>0$
observed at the Tevatron means that the top quark tends to move
along with the valence quark than to move in the opposite direction.
Second, $A^t_{\rm FB}$ depends on the collider energy
$\sqrt{s}$. We found that as $\sqrt{s}$ increases, $A^t_{\rm
FB}$ increases monotonically in the $W^\prime$ model but decreases
monotonically in the diquark model. This means that if the two
models predict a same $A^t_{\rm FB}$ at the Tevatron, then as
$\sqrt{s}$ increases to the LHC energy, the tendency of top quark to
move with the valence quark ($u$ or $d$) in the $W^\prime$ model
should be larger than that in the diquark model.

\begin{figure}[thb]
\epsfig{file=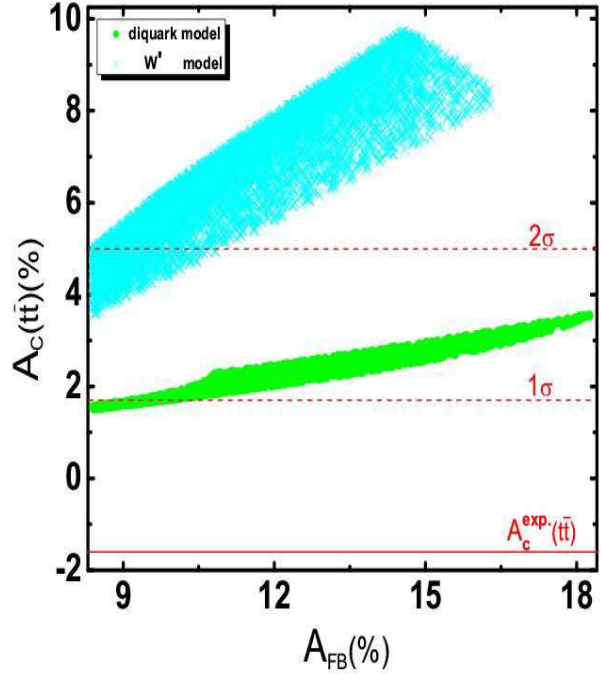,width=8cm,height=9cm} \vspace{0.3cm}
\vspace{-1cm} \caption{The correlation between $A_{\rm FB}^{t}$
at the Tevatron and $A_{C}(t\bar{t})$ at the LHC. \label{fig1}}
\end{figure}
In Fig.~\ref{fig1} we show the correlation between $A_{\rm
FB}^{t}$ at the Tevatron and $A_{C}(t\bar{t})$ at the LHC in
these two models. Such results are obtained by scanning over the
two-dimension parameter space of the models and keeping only the
samples surviving the Tevatron constraints. We see that $A_{\rm
FB}^{t}$ and $A_{C}(t\bar{t})$ are of the same sign and with the
increase of $A_{\rm FB}^{t}$ the value of $A_{C}(t\bar{t})$
increases too. This behavior can be understood by noting the
following three points. The first is that in the $t\bar{t}$ rest
frame the top and the anti-top outgo back to back. So, regardless
the underlying dynamics, we always have $|\eta_t| =
|\eta_{\bar{t}}|$. The second is that for the $t$-channel process
$d\bar{d} \to t \bar{t}$ or the $u$-channel process
$u\bar{u}\rightarrow t\bar{t}$ at $pp$ colliders like the LHC, the
$t\bar{t}$ rest frame tends to be boosted along the direction of $d$
or $u$ quark since they are the valence quarks in proton. For a
given event, the direction of the valence quarks is definite. Then,
if the scattering angle $ \theta_{tq}$ ($q=u,d$) between the
outgoing top quark and the valence quark in $t\bar{t}$ rest frame is
less (larger) than $\pi/2$, $|\eta_t|$ defined in the laboratory
frame tends to be larger (less) than $|\eta_{\bar{t}}|$. And the
last point is if the top quark has equal probability to move along
and to move in opposite to the valence quark direction at the LHC
(corresponding to $A^t_{\rm FB} = 0$ in $p\bar{p}$ collision), the
number of events with $|\eta_t|
> |\eta_{\bar{t}}|$ should be same as that with $|\eta_t| <
|\eta_{\bar{t}}|$, and hence $A_C(t\bar{t})=0 $; if the former
probability exceeds the latter probability (corresponding a positive
$A^t_{\rm FB}$ in $p\bar{p}$ collision), more events with $|\eta_t|
> |\eta_{\bar{t}}|$ than with $|\eta_t| < |\eta_{\bar{t}}|$ should
be obtained and thus $A_C(t\bar{t})$ is positive. This analysis
shows that $A^t_{\rm FB} $ at the Tevatron can be treated as an
indicator of $A_C(t\bar{t})$ at the LHC.

Fig.~\ref{fig1} also indicates that $A_C(t\bar{t})$ in the
$W^\prime$ model is usually several times larger than that in the
diquark model for a given value of $A^t_{\rm FB}$. One underlying
reason is, as we mentioned before, the probability of the top quark
to move along with the valence quark in the $W^\prime$ model exceeds
that in the diquark model. Another reason is from the parton
distribution of the initial states: at the Tevatron we have
$P_{d\bar{d}}:P_{u\bar{u}} \simeq 1:4$, while at the LHC
$P_{d\bar{d}}:P_{u\bar{u}} \simeq 1:2$. So when both models predict
a same $A^t_{\rm FB}$ at the Tevatron, the parton distribution in
the $W^\prime$ model is relatively enhanced at the LHC.

Another striking feature of Fig.~\ref{fig1} is that a large portion
of the samples in the $W^\prime$ model have been ruled out by the
measured value of $A_C(t\bar{t})$ at $2\sigma$ level, which
implies that the $W^\prime$ model has already been tightly limited
by the charge asymmetry. In contrast, in the diquark model the
$A_C(t\bar{t})$ value always lie within $2\sigma$ range of its
experimental central value. We checked that the $A_C(t\bar{t})$ value
in the diquark model will be further reduced at the LHC as $\sqrt{s}$
is raised to 14 TeV.

In getting Fig.1, we note that, since the new physics contributions
to the $t\bar{t}$ cross section are relatively small, both $A_C$ and
$A^t_{FB}$ can be approximated as the SM value plus the new physics
effect: $A_C \simeq A^{SM}_C+ \delta A_C$ and $A^{t}_{FB} \simeq
A^{t\ SM}_{FB} +\delta A^{t}_{FB}$. For the values of $A^{SM}_C$ and
$A^{t\ SM}_{FB}$, we use their NLO QCD results: $A^{\rm
SM}_{C}(t\bar{t})=0.0130$ \cite{cms-ac} and $A^{t\ SM}_{FB} = 0.038
$ (which is obtained by the MCFM package \cite{top-afb-cdf}). In
calculating $\delta A_C$ and $\delta A^t_{FB}$, we encounter two
kinds of cross sections: the SM cross sections
$\sigma^{SM}_{t\bar{t}}$ and the new physics corrections
$\delta\sigma_{t\bar{t}}$. We use the tree-level expression of
$\delta \sigma_{t\bar{t}} $ due to the absence of its high order QCD
correction in literatures, while for the $\sigma^{SM}_{t\bar{t}}$,
we use its most precise NNLO result, which is obtained by
multiplying its LO prediction by a K factor, i.e. $K\simeq 1.7$ for
the LHC \cite{top-cross-th-lhc} and $K\simeq1.3$ for the Tevatron
\cite{top-cross-th-tev}.

\begin{figure}[thb]
\epsfig{file=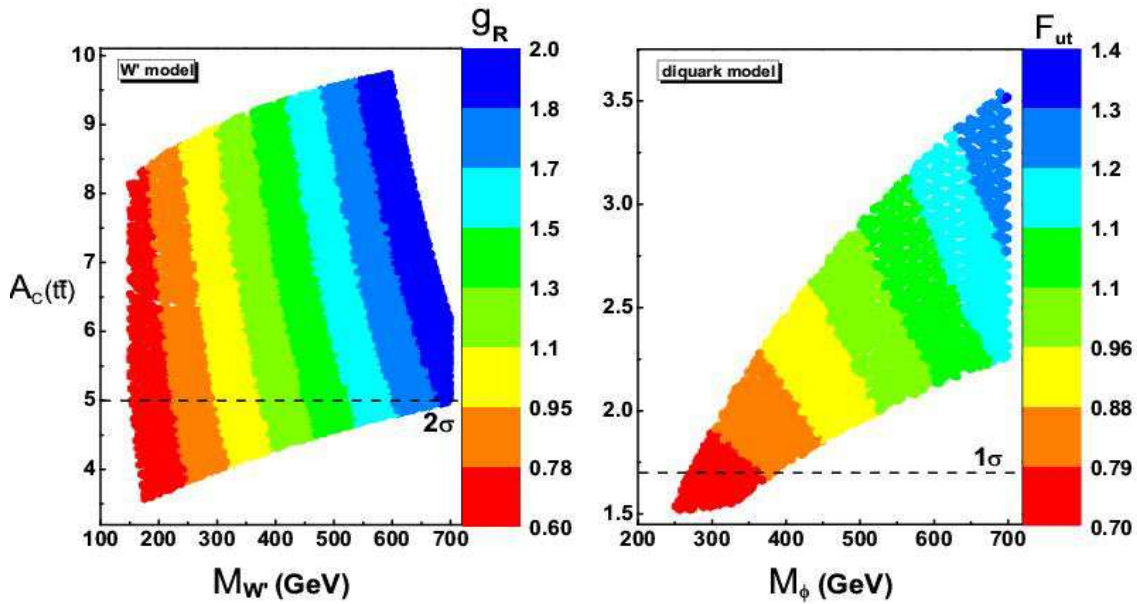,width=15cm,height=8cm} \vspace{-0.5cm}
\caption{The dependence of $A_{C}(t\bar{t})$ on the model
parameters. Samples shown here satisfy the Tevatron measurements at
$1\sigma$ level described in the text.} \label{fig2}
\end{figure}

In Fig.~\ref{fig2} we show the dependence of $A_{C}(t\bar{t})$
on the model parameters such as the coupling strength and the new
particle mass. Due to the difference in kinematic features of the
$t$ and the $u$ channels, the mass ranges favored by $A^t_{FB}$
and $\sigma(t\bar{t})$ are $ 150{\rm GeV} < m_{W^\prime} < 700
{\rm GeV}$ and $ 250{\rm GeV} < m_{\phi} < 700 {\rm GeV}$ for the
two models respectively. This figure indicates that for a given new
particle mass the coupling coefficient ($f_{ut}$ or $g_R$) is
restricted in a certain region, and as the new particle becomes
heavy, the region moves upward. This is because we have required the
samples shown in the figure to explain the $A^t_{\rm FB}$ anomaly
and at same time to satisfy the $\sigma_{t\bar{t}}$ constraint. This
figure also indicates that a heavy new particle along with a strong
coupling can predict a large $A_{C}(t\bar{t})$. We checked this case
and found it usually corresponds to a large $A^t_{\rm FB}$ at the
Tevatron.

\begin{figure}[htbp]
\includegraphics[width=2.5in,height=3.2in]{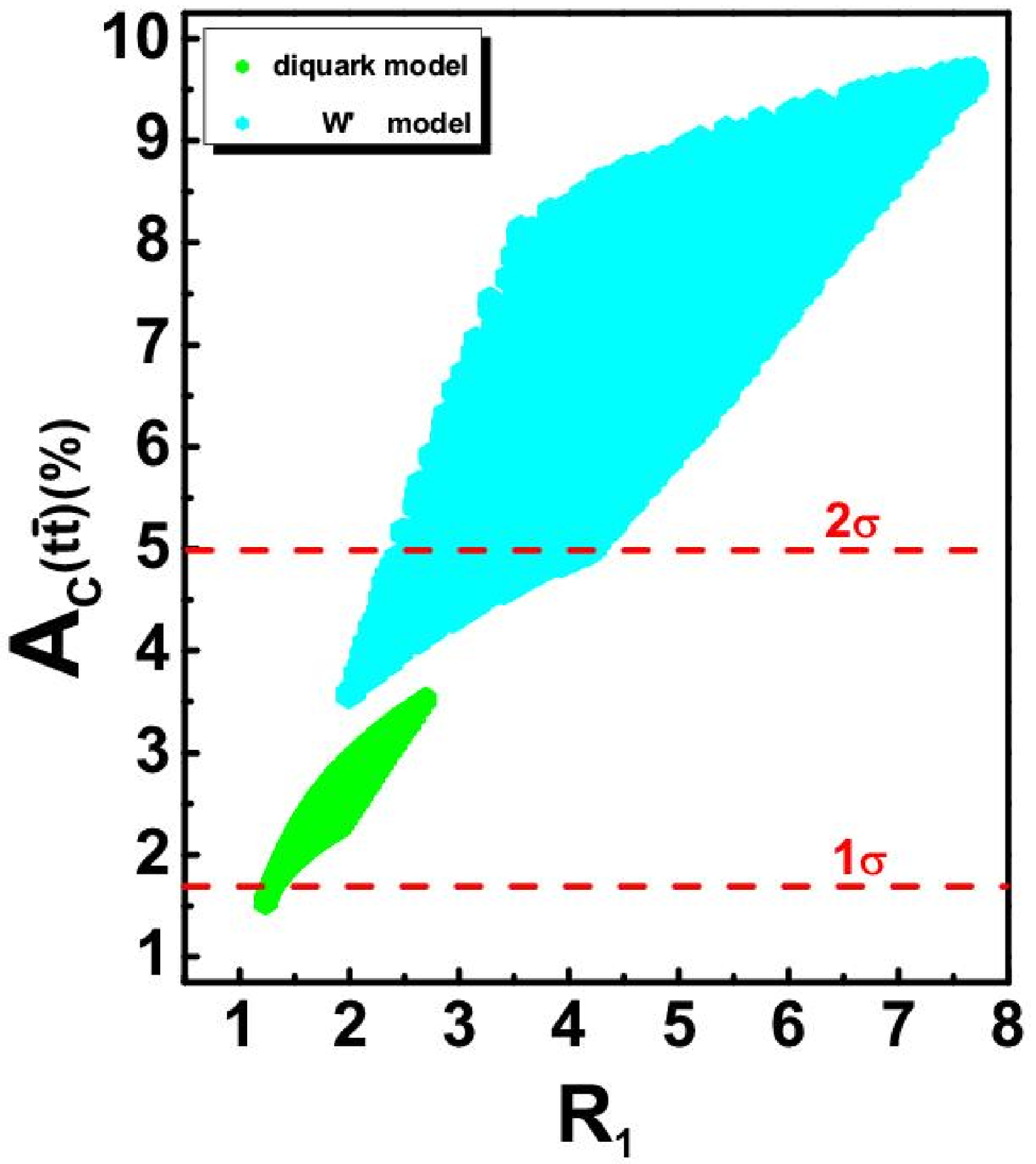}%
\hspace{-0.02in}%
\includegraphics[width=2.5in,height=3.2in]{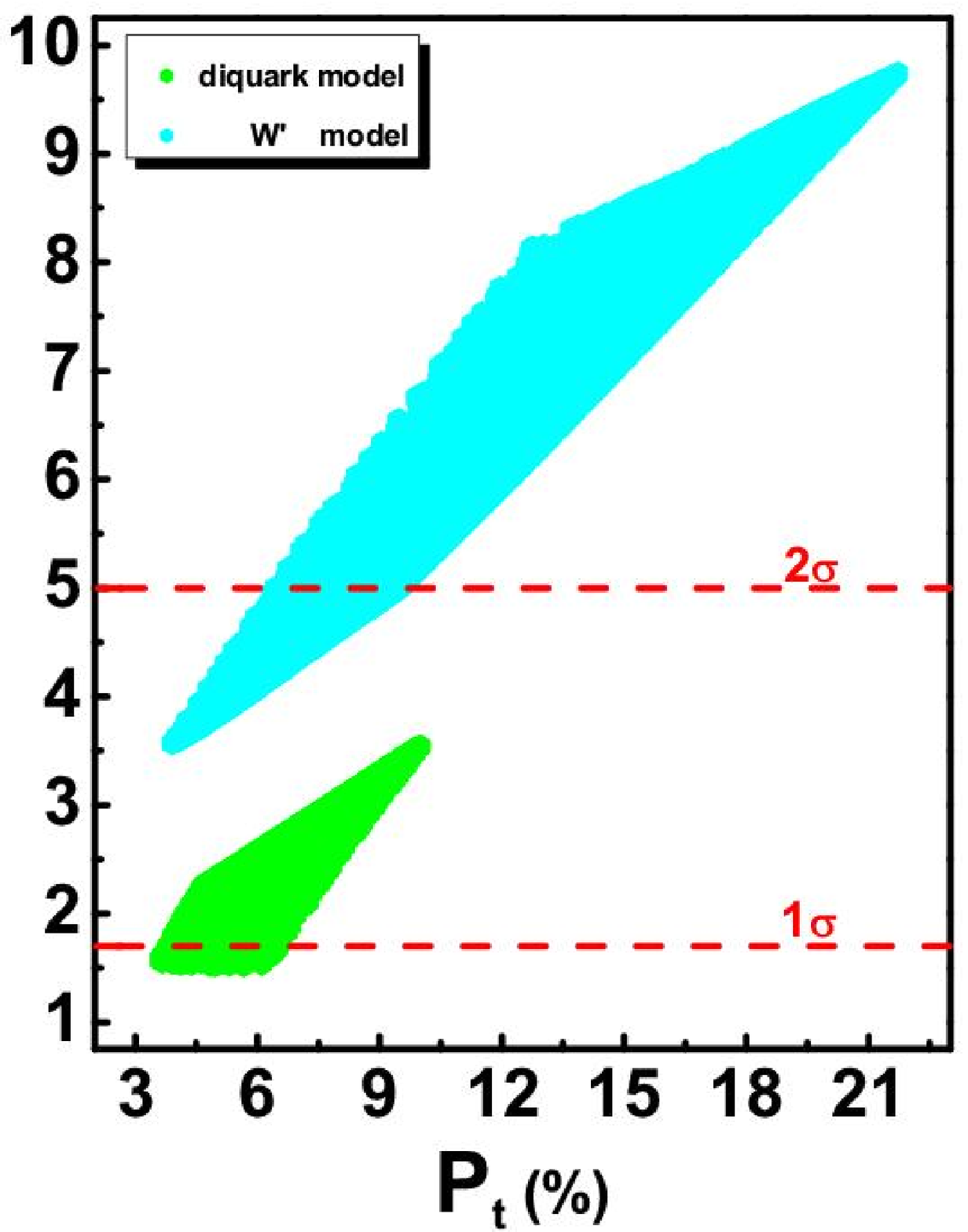}%
\hspace{0in}%
\vspace{-0.5cm} \caption{The correlations of $A_{C}(t\bar{t})$
with $R_1$ and $P_t$ at the LHC respectively. \label{fig3}}
\end{figure}

In the left frame of Fig.~\ref{fig3} we show the correlation of
the $A_{C}(t\bar{t})$ with the ratio $R_1$ defined by
Eq.~(\ref{R1}). As we mentioned before, for the $t$-channel or the
$u$-channel $t\bar{t}$ production, the Rutherford singularity
tends to push more events to high $M_{t\bar{t}}$ region so that
$R_1$ may be significantly larger than unity. This is reflected in
the $W^\prime$ model where $R_1$ is in the range of 2.0 and 7.7
and in the diquark model where $R_1$ varies from 1.2 to 2.7. Since
the predicted $R_1$ is in two separated regions, $R_1$ may be
utilized to discriminate the models. We checked the reason for the
difference and found that the cancellation between the pure new
physics contribution and the interference contribution in the
$W^\prime$ model is not as strong as that in the diquark model. We
also note that the LHC with higher luminosity is capable of
exploring the models with $R_1 > 2$ \cite{inv-mass-lhc}. So we
conclude that the quantity $R_1$ is complementary to
$A_C(t\bar{t})$ in testing the models.

Since the new interactions violate parity and hence can lead to top
quark polarization asymmetry $P_t$ at the LHC, in the right frame of
Fig.~\ref{fig3} we show the correlation of $A_{C}(t\bar{t})$
with $P_t$. This figure indicates that the value of $P_t$
increases with the increase of $A_{C}(t\bar{t})$ with its
maximum value reaching $22\%$ and $10\%$ for the two models
respectively. To roughly estimate the observability of such
asymmetry, we calculate the statistical significance $N_S$ defined
in \cite{pt-np1} for an integrated luminosity of 1 fb$^{-1}$ without
considering the cut efficiency and the systematic uncertainties. We
find that for nearly all the samples in the models, the predicted
$P_t$ can reach its $3\sigma$ sensitivity, which is $1.20\%$ for the
$W'$ model and $2.15\%$ for the diquark model.

\subsection{Observables in single top production }
\begin{figure}[htb]
\epsfig{file=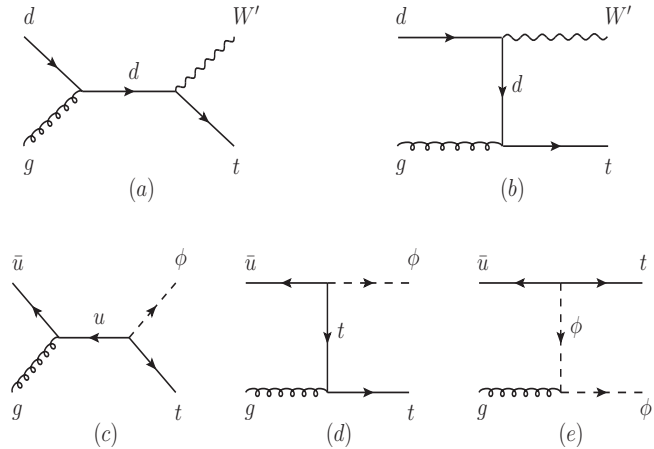,width=9cm,height=6cm} \vspace{-0.5cm}
\caption{Feynman diagrams contributing to the single top productions
at the LHC} \label{fig4}
\end{figure}

In the $W^\prime$ (diquark) model, the associated production of
single top quark with $W^\prime$ ($\phi$) proceeds by the Feynman
diagrams shown in Fig.~\ref{fig4}. The total production rates (top
events plus anti-top events) can reach 60 pb and 160 pb for the
surviving samples in the two models respectively.

\begin{figure}[htbp]
\includegraphics[width=2.5in,height=3.2in]{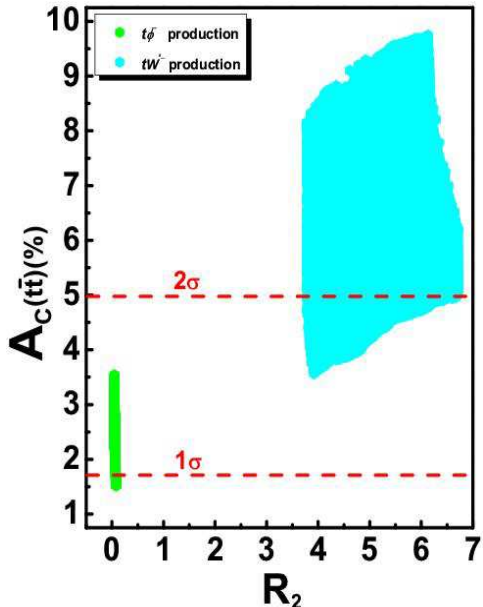}%
\hspace{0.05in}%
\vspace{-0.8cm} \caption{The correlations between the
$A_{C}(t\bar{t})$ and $R_2$ at the LHC.} \label{fig5}
\end{figure}

Due to the electric charge carried by $W^{\prime -}$ ($\phi^-$), the
production rates of the top quark and anti-top quark are not equal.
Since the initial state is $d g$ ($\bar{u}g$) for the top production
and $\bar{d} g$ ($u g$) for the  anti-top production, the parton
distributions determine $R_2 > 1 $ for the $W^\prime$ model and $R_2
< 1$ for the diquark model, where $R_2$ denotes the charge asymmetry
of the associated production defined in Eq.~(\ref{R2}). From
Fig.~\ref{fig5}, we find $3.6 <R_2 < 6.8 $ in the $W^\prime$ model
while $R_2 < 0.2 $ in the diquark model. In our calculation we also
find that, although the rate of the $t W^{\prime -}$ production
decreases monotonically as $W^\prime$ becomes heavy, the ratio
$R_{2}$ increases. The reason is that the distribution function of
the sea quark $\bar{d}$ is more suppressed in high proton momentum
fraction region.

In order to further test two models, we investigate the kinematical
distributions of the single top productions. As an illustration, we
take the best point for each model. The
best point is determined by minimizing the $\chi^2$ function
defined as
\begin{eqnarray}
\chi^2= \sum_i\frac{(O_i^{\text{theory}} -
O_i^{\text{measured}})^2}{\sigma_i^2},
\end{eqnarray}
where the observables $O_i$ are $A^t_{\rm FB}$ and
$\sigma(t\bar{t})$ at the Tevatron and $A_{C}(t\bar{t})$ at the LHC.
We add the experimental and the SM errors in quadrature to
calculate $\sigma_i$. For the $W^\prime$ model the best point is
found to be at $g_R=0.605$ and $m_{W^\prime}= 697.85$ GeV, with
$\chi^2/{\rm dof}=4.69/3$; while for diquark model the best point is
at $f_{ut}=0.91$ and $m_\phi=442.43$ GeV, with $\chi^2/{\rm
dof}=1.47/3$. In Table I we present the predictions for the
observables at the best points.

\begin{table} \tabcolsep0.1in
\caption{Predictions of the $W^\prime$ model and the diquark model
at the best point. $X$ denotes $W'$ or $\phi$.
New physics contributions to the cross sections
at the Tevatron (LHC) are in unit of fb (pb).}
\small
\begin{tabular}{|c|c|c|c|c|c|c|c|c|c|c|}
\hline
  & \multicolumn{2}{|c|}{Tevatron}  & \multicolumn{8}{|c|}{LHC}\\
\cline{2-11}
 &$\Delta\sigma(t\bar{t})$ & $A^{t}_{\rm FB}$ & $\Delta\sigma(t\bar{t})$ & $A_{c}(t\bar{t})$
& $P_t$ & $A_{C}(XX)$ & $R_{1}$ & $R_{2}$&$\sigma(tX)$& $\sigma(XX)$ \\
\hline
$W^{\prime}$ & 107.84 & 0.054 & -0.71  & 0.011 & -0.006 & $0.05$ & 0.09 & 6.7&0.26 &0.002\\
\hline
diquark  & 831.20 & 0.120 & 0.99 & 0.021 & 0.055  &  $-0.69$  & 1.54 & 0.06 &2.5& 0.87 \\
\hline
\end{tabular}
\normalsize
\end{table}

In our analysis we assume $W^{\prime -}$ and $\phi^-$ mainly decay
as $W^{\prime -} \to \bar{t} d$ and $\phi^- \to \bar{t} \bar{u}$
with the anti-top quark decaying hadronically so that $W^\prime$
and $\phi$ can be reconstructed. In this way, the associated
productions may be disentangled from the $t\bar{t}$ production
\cite{wp} which acts as the main background. Using the
MadGraph5/MadEvent \cite{mdme}, we study the signal $3j+2b+l+
\rlap{$\,/$}E_T$ at the parton level under the basic cuts at the
LHC, where $\rlap{$\,/$}E_T$ denotes the missing transverse energy.

\begin{figure}[htbp]
\includegraphics[width=3in,height=3.5in]{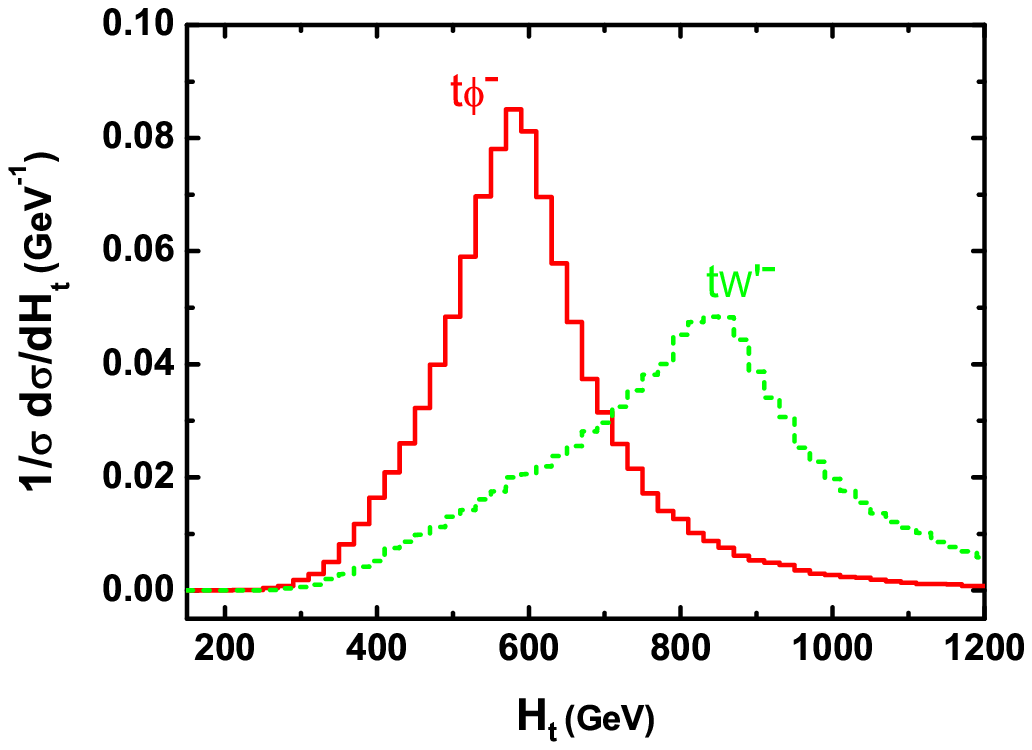}%
\hspace{0in}%
\includegraphics[width=3in,height=3.5in]{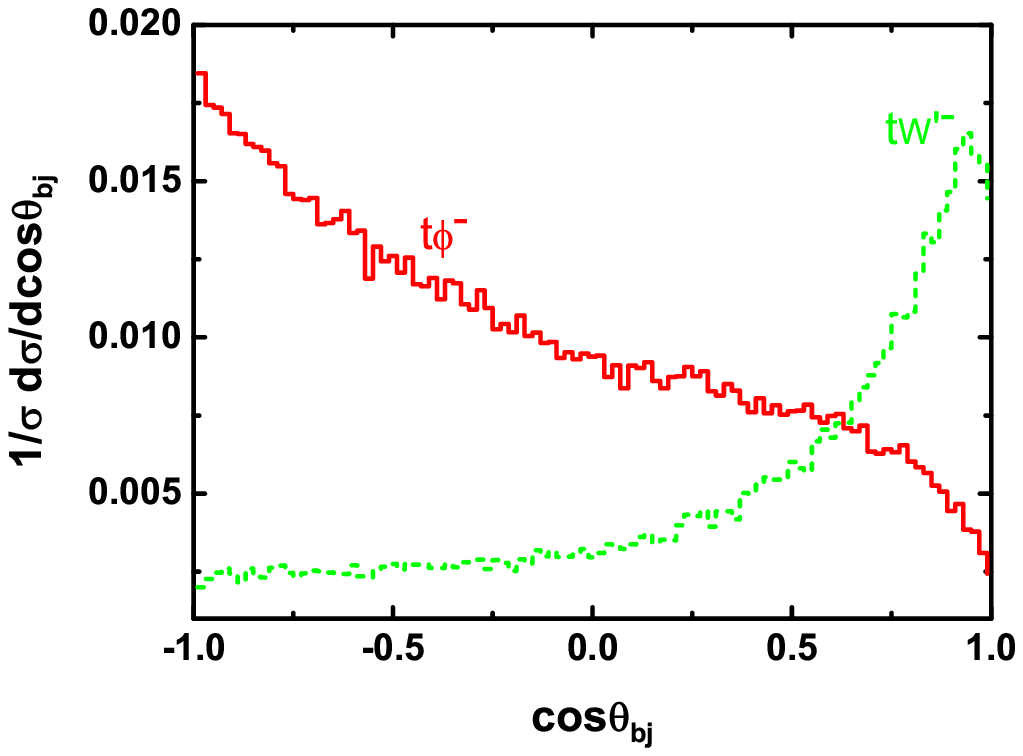}%
\hspace{0in}%
\vspace{-1.0cm} \caption{The distributions of $H_t$ and
$\cos\theta_{bj}$ in the single top productions at the LHC. Here the
$b$-jet and the light jet are required from same new particle.}
\end{figure}

In Fig.~6 we display the distributions of the total transverse
energy $H_{T}$ and the angle between the $b$-jet and the light jet
coming from $W'$($\phi$), which are all defined in the
laboratory frame. The left panel of this figure shows that the most
events from $t W^\prime$ have lower $H_{T}$ than those from $t
\phi^{-}$. The reason is that in the considered case $W^\prime$
is lighter than the diquark state. The right panel shows that the
$b$-jet is inclined to fly along the light jet in the $W'$ model,
while to fly in opposite to the light jet in the diquark model. This
is because, although the decay products of $W^{\prime}$($\phi$) are
boosted along the direction of $W^{\prime}$($\phi$), the massive
anti-top from the $W^{\prime}$($\phi$) decay may kick its $b$-jet in
certain direction so that the $b$-jet can deviate from the boost
direction. Actually, we find that the $b$-jet from a left-handed
anti-top quark (as in the $W^{\prime}$ model) tends to fly along the
direction of the anti-top quark \cite{han}, which is also the
direction of the light jet from the $W^{\prime}$ decay; while the
$b$-jet from a right-handed anti-top quark (as in the case in the
diquark model) tends to fly in the opposite direction.

For the charge asymmetry in single top production, due to the large
jet multiplicities and moderate $b$-tagging efficiency in the
process, the measurement will be somewhat challenging at the LHC.
However, we noted that the peak values of $H_{T}$ ($>500$GeV) in
both models are much larger than that in the SM ($\sim 350$GeV).
With higher luminosity and higher kinematic cuts, the measurements
of the differential cross sections and the single top charge
asymmetries versus $H_{T}$ will be useful to discover the signals
\cite{single-top-ac1}. Moreover, the $b$-jet angular distribution
may serve as a complementary discriminator for the background, since
the distribution of $\cos\theta_{bj}$ in the SM is relatively flat
in comparison with the signals. The detailed analysis of the
backgrounds depends on the full detector simulation which is
partially studied in Ref. \cite{top-resonances}.

\subsection{Observables in $W'^+W'^-$ and $\phi^+\phi^-$ productions}
\begin{figure}[htbp]
\includegraphics[width=4.2in,height=3.2in]{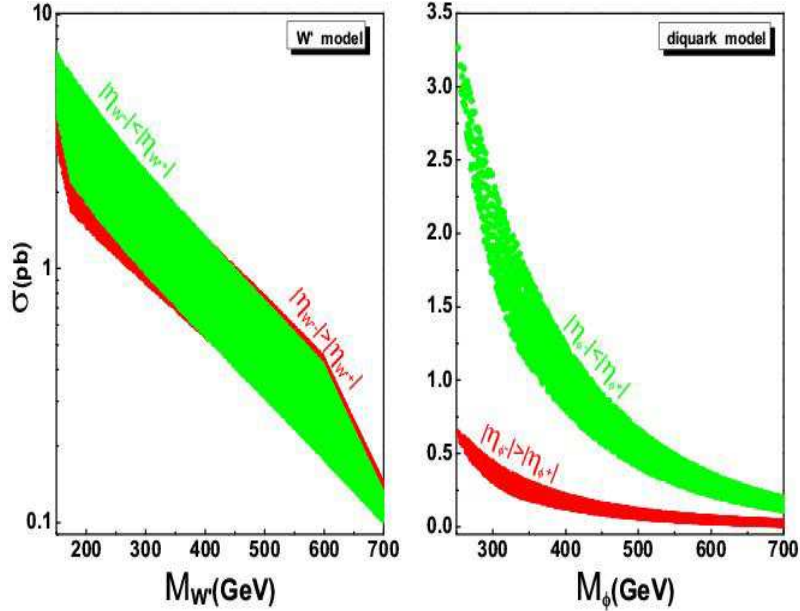}%
\hspace{0in}%
\vspace{-0.8cm} \caption{Pair production rate at the LHC versus the
corresponding particle mass.} \label{fig7}
\end{figure}

Due to the interactions introduced in Sec.~II, the $W'^+W'^-$
production proceeds only by the parton process $d\bar{d} \to
W^{\prime +} W^{\prime -}$ through exchanging a top quark, while the
 $\phi^+\phi^-$ production may proceed either by $u\bar{u} \to \phi^+
\phi^-$ or by $g g \to \phi^+ \phi^-$ (via $gg\phi\phi$ and
$g\phi\phi$ interactions). We checked our results for the
$\phi^+\phi^-$ production and found that the gluon annihilation
contribution is usually negligibly small. One main reason is that
for the surviving samples presented in Fig.~\ref{fig2}, $\phi$ is
usually heavy and thus the gluon distribution in proton is
suppressed.  We also found that, for given $m_{W^\prime} = m_\phi =
m_P$, the $\phi^+\phi^-$ production rate is slightly lower than the
$W'^+W'^-$ rate. This is shown in Fig.~\ref{fig7}, where one can
learn that for $m_P = 250$ GeV, $\sigma (W^{\prime +} W^{\prime -})$
may exceed 6 pb while $\sigma (\phi^+ \phi^-)$ can only reach 4 pb.

\begin{figure}[htb]
\epsfig{file=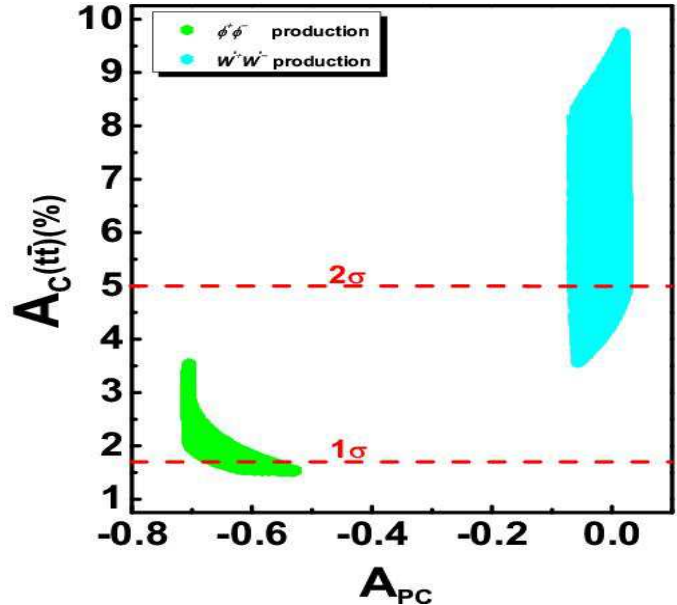,width=9cm,height=8cm} \vspace{-0.5cm}
\caption{The correlation of $A_{C}(t\bar{t})$ with
 $A_{C}(W'^+W'^-)$ and $A_{C}(\phi^+\phi^-)$ at the LHC respectively.} \label{fig8}
\end{figure}

Although the pair production rates are moderate at the LHC with
$\sqrt{s} = 7$ TeV, the charge asymmetry $A_{C}$ can still be
sizable because it only reflects the unbalance between the particle
and its charge conjugate state in boosting along the valence quarks.
In Fig.~\ref{fig8} we show the charge asymmetry $A_{C}$ in the two
models. This figure indicates that in the $W^\prime$ model the
$A_{C}(W'^+W'^-)$ fluctuates around zero, while in the diquark model
$A_{C}(\phi^+\phi^-)$ varies between $-0.5$ and $-0.8$. These
results can be understood from Fig.~\ref{fig7}, which shows that for
$m_{W^\prime} < 408$ GeV the cross section with
$|\eta_{W^{\prime-}}|< |\eta_{W^{\prime+}}|$ is slightly larger than
that with $|\eta_{W^{\prime-}}| > |\eta_{W^{\prime+}}|$, and with
the increase of $m_{W^\prime}$ this relation is reversed; while in
the diquark model the corresponding former rate is always larger
than the latter rate to obtain a significant negative
$A_{C}(\phi^+\phi^-)$.

We note that in the SM the value of $A_{C}$ for the $W^-W^+$
production is positive, while in the $W^\prime$ model  the value of
$A_{C}(W'^+W'^-)$ is negative for a light $W^\prime$. The difference
comes from the masses of mediators. In the SM, the main contribution
to the $W^+W^-$ production is through the $t$ channel by mediating a
massless light quark, while in the $W^{\prime}$ model, it is top
quark that mediates the process of the $W'^+W'^-$ production. We
checked that if we set $m_t$ to zero, $A_{C}$ in $W^\prime$ pair
production will become positive as $A_{C}(W^+W^-)$ in the SM. We
also note that in the diquark model, even with the constraints from
$A_C(t\bar{t})$, the value of $A_{C}(\phi^+\phi^-)$ can still
deviate significantly from zero. We checked that at the LHC with
$\sqrt{s}=14$ TeV the rates for these productions are usually
enhanced by about $3\sim4$ times, while $A_{C}$ changes little in
both models.

\section{Conclusion}
In this paper we discussed the potential of the LHC to discriminate
the $W'$ model and the diquark model which were used to explain the
$A^t_{\rm FB}$ anomaly measured at the Tevatron. With the
constraints from the Tevatron, we examine the charge and
polarization asymmetries in $t\bar{t}$ production, the charge
asymmetries in single top production and $W^\prime$$(\phi)$ pair
production at the LHC with $\sqrt{s}=7$ TeV. We found that the
predictions of these observables may be large enough to reach their
detectable levels at the LHC. In particularly, the recent
measurement of the charge asymmetry in $t\bar{t}$ production from
the LHC has already imposed a strong limit on the $W^\prime$
explanation of the $A^t_{\rm FB}$ anomaly. We also found that each
observable in the two models shows different characteristics and a
joint analysis of these observables at the LHC can help to
discriminate the two models.

\section*{Acknowledgement}

Lei Wu thanks Fabio Maltoni and Johan Alwall for
helpful discussion of Madgraph. This work was supported in part by
HASTIT under grant No. 2009HASTIT004, by the National Natural
Science Foundation of China (NNSFC) under grant Nos. 10821504,
10725526, 10775039, 11075045, by the Project of Knowledge
Innovation Program (PKIP) of Chinese Academy of Sciences under grant
No. KJCX2.YW.W10, and by the
Grant-in-Aid for Scientific Research (No.~14046201) from Japan.

\appendix
\section{An asymmetric left-right model with a light $W^{\prime}$}

The asymmetric left-right model with light $W^{\prime}$ was proposed
in \cite{Asy-LR,mfv}. It is based on the gauge group $SU(2)_L
\bigotimes SU(2)_R \bigotimes U^\prime (1)$ and assumes that only
the first and third generation right-handed quarks transform
nontrivially under the group $SU(2)_R$\cite{mfv}. The symmetry
breaking starts with $SU(2)_R \bigotimes U^\prime(1) \rightarrow
U(1)_Y $ to obtain the SM hypercharge $Y = 2 T_3^R + Y^\prime$, and
subsequently $SU(2)_L \bigotimes U(1)_Y \rightarrow U(1)_{EM} $ to
obtain $Q = T_3^L + Y/2$. For the first breaking, a $SU(2)_R$
triplet Higgs field is introduced so that the neutral gauge bosons
$Z^\prime$ of the $SU(2)_R$ group is significantly heavier than the
charged boson $W'$\cite{mfv,Asy-LR}. Two distinctive features of the
model are exhibited in \cite{mfv}. One is, after choosing specific
rotation matrices to transform right-handed quarks from flavor basis
to mass eigenstates, $W^\prime$ may couple to flavors in the
combination $(t,d)_R$ with unsuppressed strength, while $Z^\prime$
only has flavor conserving interactions, i.e.
\begin{eqnarray}
\mathcal {L} &=& g_R \bar{t}\gamma^{\mu} P_{R} d W^{\prime}_{\mu} +
\sum_{q_i=u,t} \{ \bar{q}_i \gamma^\mu ( g_{Li} P_L + g_{Ri} P_R )
q_i  \} Z^\prime_\mu + \hbox{h.c.}\;. \label{wprime}
\end{eqnarray}
Such specific choice, as shown in \cite{mfv}, is
phenomenologically favored by several anomalies in top physics and
B physics observed at the Tevatron. The second feature is, unlike
the traditional flavor universal left-right model where the quarks
acquire masses by interacting with $SU(2)_L \bigotimes SU(2)_R$
bi-doublet fields\cite{Frank}, the quark masses are generated in a
complex way. For example, the first and third generation
right-handed quarks may have Higgs terms like
\begin{eqnarray}
\frac{<\phi_R> f^{d}_{ij}}{M} \frac{1}{<\phi_R>}{\bar{q'}}_R^i
\phi_R^\dagger H_L {q'}_L^{j} + \frac{<\phi_R>
f^{u}_{ij}}{M}\frac{1}{<\phi_R>}{\bar{q'}}_R^i
\tilde{\phi}_R^\dagger \tilde{H}_L {q'}_L^{j} , \label{operator}
\end{eqnarray}
where flavor indices $i$ and $j$ are $i = 1,3$ and $j = 1,2,3$,
$\phi_R$ and $H_L$ are doublet fields under the group $SU(2)_R$ and
$SU(2)_L$ respectively with $\tilde{\phi}_R ^ a
=\epsilon_{ab}\phi^{*b}_R $ and
$\tilde{H}_L^a=\epsilon_{ab}H_L^{*b}$, and $<\phi_R>$ denotes the
vacuum expectation value (vev) of the neutral component of $\phi_R$;
whereas the second generation right-handed quarks take on the more
conventional form
\begin{equation}
f^d_j{\bar{q'}}_R^2  H_L {q'}_L^{j} + f^u_j{\bar{q'}}_R^2
\tilde{H}_L {q'}_L^{j}.
\end{equation}
Obviously, once the field $\phi_R$ gets its vev the SM mechanism for
mass generation is recovered  with the quark Yukawa coupling
coefficients $Y_{ij}$ given by $\frac{<\phi_R> f_{ij}}{M} $ for $i =
1,3$ and $f_j$ for $i=2$. In addition, as suggested by \cite{mfv},
the five dimension operators in Eq.(\ref{operator}) may be generated
by integrating out heavy $SU(2)_{L,R}$-singlet fermions with mass
scale $M$, which usually carry appropriate hypercharge.

In the $W^\prime$ model, the additional contribution to the
$t\bar{t}$ production comes from the $t$-channel process $q\bar{q}
\to t \bar{t}$ via the exchange of $W^\prime$ or neutral/charged
component fields of the $\phi_R$. Obviously, if the component fields
are heavy and/or if  $<\phi_R>$ is much larger than the electro-weak
breaking scale so that the $\bar{q} q^\prime \phi_R$  interactions
are suppressed (see Eq.\ref{operator}), the latter contribution can
be safely neglected, which was done in literature \cite{Asy-LR,mfv}.

\end{document}